\documentclass[A4]{article}

\usepackage{color}
\usepackage{graphicx}

\font\scap=cmcsc10

\newcount\eqnumber
\def\neweq{{\rm{(\the\eqnumber)}}\global\advance\eqnumber by 1}
\def\eqdef#1{\eqno\xdef#1{\the\eqnumber}\neweq}
\def\newaeq{{\rm{(\the\eqnumber a)}}\global\advance\eqnumber by 1}
\def\eqdaf#1{\eqno\xdef#1{\the\eqnumber}\newaeq}
\def\eqdisp#1{\xdef#1{\the\eqnumber}\neweq}
\def\eqdasp#1{\xdef#1{\the\eqnumber}\newaeq}

\newcount\refnumber
\def\newref{{\the\refnumber}\global\advance\refnumber by 1}
\def\refdef#1{{\xdef#1{\the\refnumber}}\newref}

\newcount\fignumber
\def\newfig{{\the\fignumber}\global\advance\fignumber by 1}
\def\figdef#1{{\xdef#1{\the\fignumber}}\newfig}

\font\tenmsb=msbm10
\font\sevenmsb=msbm7
\font\fivemsb=msbm5
\newfam\msbfam
\textfont\msbfam=\tenmsb
\scriptfont\msbfam=\sevenmsb
\scriptscriptfont\msbfam=\fivemsb

\def\smallskip{\vskip 3pt}
\def\medskip{\vskip 6pt}
\def\bigskip{\vskip 12pt}

\begin{document}

\centerline{\bf On the singularity structure of a discrete modified-Korteweg-deVries equation}
\bigskip

\medskip{\scap Basil  Grammaticos}\quad
{\sl  Universit\'e Paris-Saclay, CNRS/IN2P3, IJCLab, 91405 Orsay, France} 
and {\sl  Universit\'e de Paris, IJCLab, 91405 Orsay, France}
\medskip{\scap Thamizharasi Tamizhmani}
\quad{\sl  SAS, Vellore Institute of Technology, Vellore - 632014, Tamil Nadu, India}
\medskip {\scap Ralph Willox} \quad
{\sl Graduate School of Mathematical Sciences, the University of Tokyo, 3-8-1 Komaba, Meguro-ku, 153-8914 Tokyo, Japan }

\bigskip
{\sl Abstract}
\smallskip
We study the singularities of a modified lattice Korteweg-deVries (KdV) equation and show that it admits three families of singularities, with analogous properties to those found in the lattice KdV equation. 
The first family consists of localised singularities which can occupy an arbitrarily large domain but which are, nevertheless, always confined. The second family consists of one or more lines extending all the way from the south-west to the north-east on the plane, involving a single finite value that depends on the parameter that appears in the equation. We argue that the infinite extent of this singularity is not incompatible with the confinement property or with  the integrability of the equation. The third family consists of horizontal strips in which the product of values on vertically adjacent lattice sites is equal to 1. In the case of the lattice KdV equation this type of singularity was dubbed `taishi'. The taishi for the modified lattice KdV equation can interact with singularities of the other two families, giving rise to very rich {and quite intricate} singularity structures. Nonetheless, these interactions can be described in a compact way through the formulation of a symbolic representation of the dynamics that is similar to, but in a sense, simpler than that for the KdV case. We give an interpretation of this symbolic representation in terms of a Box\&Ball system related to the ultradiscrete mKdV equation. These results show that taishi-type singularities are not limited to the lattice KdV equation, but might very well be a general feature of integrable lattice equations with deep connections to other facets of their integrability.

\bigskip
1. {\scap Introduction}
\medskip
Singularities play a major role in physics and mathematics [\refdef\goriely]. As M. Kruskal used to point out, `singularities are where interesting things happen'. In the domain of mathematical modelling of  physical systems the study of singularities turned out to be a most useful tool. Physical phenomena can display, depending on conditions, either regular or irregular behaviour. One thus expects that their mathematical models would exhibit the same diversity of behaviour. This is indeed the case and the prognosis of regular or irregular behaviour in dynamical systems is thus of particular interest. The theory of chaos was in fact developed to describe a particular type of irregular motion {that} precludes long-term predictions, often exhibiting a crucial dependence on the details of the initial conditions. Regular motion on the other hand is associated with predictability, allowing for long-term predictions about the evolution of a given system.

Singularities are particularly effective in the prediction of the behaviour of a given dynamical system [\refdef\physrep]. In the case of differential equations, the theory of Painlev\'e [\refdef\ars] links the singularity structure of a system to its integrable (or non-integrable) character; integrability guaranteeing regular and predictable behaviour. In the domain of discrete systems it turned out that singularities are also important. The singularity confinement conjecture [\refdef\sincon] postulated that discrete equations integrable through spectral methods must possess so-called confined singularities, where `confinement' means that any spontaneously appearing singularity again disappears after a finite number of iteration steps. 

The property of confinement of singularities was first observed  [\sincon]  in a study of the discrete (or {\sl lattice}) KdV (d-KdV) equation [\refdef\hirota]
$$x_{m+1,n+1}=x_{m,n}+{1\over x_{m+1,n}}-{1\over x_{m,n+1}}.\eqdef\zena$$

Figure \figdef\one\ shows two examples of confined singularities for the d-KdV equation. 
\bigskip
\centerline{\includegraphics[width=7cm,keepaspectratio]{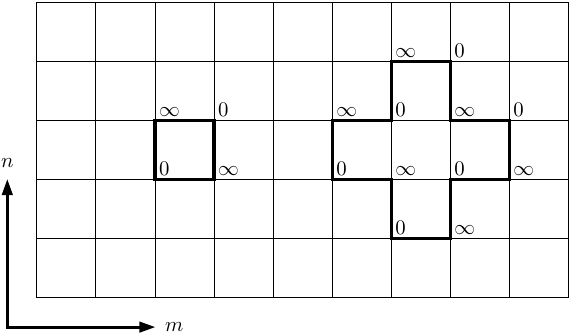}}
\smallskip\ \centerline{{Figure \one   }}
\medskip
Although the d-KdV equation was instrumental in the discovery of the singularity confinement property, the complete study of its singularities had to wait for more than a quarter century. In fact, it was not until 2021 that some of the present authors produced an exhaustive study [\refdef\doyong, \refdef\kdv] of the singularities of d-KdV and their interactions. 

Among these singularities there are several that correspond to patterns of infinite extent, as shown in Figures \figdef\two\ and \figdef\three. 
\bigskip
\centerline{\includegraphics[width=6.5cm,keepaspectratio]{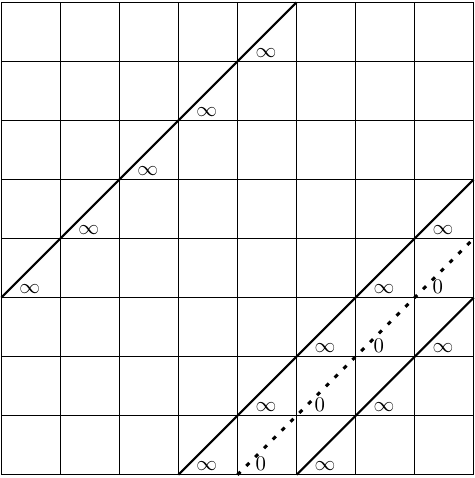}}
\smallskip\ \centerline{{Figure \two   }}
\medskip
It is important to realize however that the existence of singularities of infinite extent, such as those in Figure 2, does not invalidate the confinement property as a requirement for integrability since it is impossible to enter such a singularity coming from finite, regular, initial conditions.

The singularity shown in Figure \three\ is of a completely new type, only discovered recently [\kdv]. Its origin becomes clear if we rewrite (\zena) as
$${x_{m+1,n+1}x_{m+1,n}-1\over x_{m+1,n}}={x_{m,n}x_{m,n+1}-1\over x_{m,n+1}}.\eqdef\zdyo$$
If at some point $(m,n)$ on the lattice the product $x_{m,n}x_{m,n+1}$ happens to be equal to 1 then it will remain equal to 1 at the next point to the right, i.e. at $(m+1,n)$. This singularity therefore consists in a horizontal strip, running from west to east on the lattice, where at vertically adjacent lattice points we have two values that are reciprocal. The corresponding singularity pattern is shown in the upper part of Figure \three.
\bigskip
\centerline{\includegraphics[width=6.5cm,keepaspectratio]{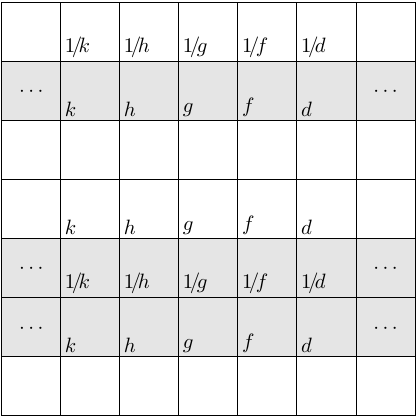}}
\smallskip\ \centerline{{Figure \three   }}
\medskip
One can also imagine situations where more than one such strip exists, some of them even being adjacent. In the lower part of Figure \three\ we show such a `double' strip, i.e. one where we have $x_{m,n}x_{m,n\pm1}=1$. Let us remark here that we could have reorganised equation (\zena) in a different way showing the existence of vertical strips that satisfy the condition $x_{m,n}x_{m+1,n}=-1$.  In [\kdv] we decided to dub the singularity associated to a strip in which the product of two adjacent values is equal to 1 (or $-1$) a `taishi'. The term taishi derives from the Japanese naming of elementary particles (which always have the suffix -shi) and the character for `band' or `strip' (`tai'). Had the taishi been isolated singularities they would have been of trifling importance. However, as we showed in [\kdv] the taishi interact with all the other singularities and these interactions lead to very interesting results. 

We shall illustrate this by an example, but before proceeding any further it is useful to introduce a handy notation. In the following, when we write 0 or $\infty$, what we really mean is that we introduce initial conditions in which a small quantity appears, say $\epsilon$, for which we take the limit $\epsilon\to0$. Thus we take any zeros to be proportional to some power of $\epsilon$, infinities to be proportional to some power of $1/\epsilon$ and the values appearing in the taishi to be such that their product is equal to $1+{\cal O}(\epsilon^p)$ for some positive integer $p$.  As explained in [\kdv] the unfettered richness of the singularities of d-KdV is only apparent if one allows different singularities to correspond to different orders in $\epsilon$. To simplify the notation we shall represent the different orders in $\epsilon$ or $1/\epsilon$ as exponents $0^1, 0^2, 0^3,\cdots$ and $\infty^1, \infty^2, \infty^3,\cdots$. In the case of the taishi, we shall keep the $\epsilon$, and the notation $\epsilon^p$ means that the product leading to the taishi is of order $1+{\cal O}(\epsilon^p)$. In what follows we shall refer to these exponents as `weights'.

In the light of this remark, the triple of lines in Figure \two\ is to be understood as representing the forward and backward evolution, according to (\zena), of an initial condition that contains the singular part $\{\infty^1,0^1,\infty^1\}$. (All other initial conditions are taken to be generic and regular and can be given either on a staircase that includes the aforementioned singular part or, for example, on the half-axes shown in Figure \figdef\four). 

One may thus wonder what would happen is one started from an initial condition $\{\infty^1,0^2,\infty^1\}$ instead. This situation is depicted in Figure \four. Looking at the patterns of 0 and $\infty$ in the (slightly greyed) right-upper part of Figure \four, which depicts the forward evolution of the initial data, one has the impression that a non-confined singularity pattern arises from the leftmost infinity, whereas the middle zero and rightmost infinity simply disappear. However this is not the case. Close inspection of the values of $x_{m+3,n}$, $x_{m+3,n+1}$ and $x_{m+3,n+2}$ shows that  their products $x_{m+3,n}x_{m+3,n+1}$ and $x_{m+3,n+1}x_{m+3,n+2}$ are indeed of order $1+{\cal O}(\epsilon)$ such that they take (finite) reciprocal values at the limit $\epsilon\to0$. This then indicates the existence of an outgoing taishi. In order to find the incoming branch of this taishi we must iterate backwards from the initial conditions. 
\bigskip
\centerline{\includegraphics[width=6.5cm,keepaspectratio]{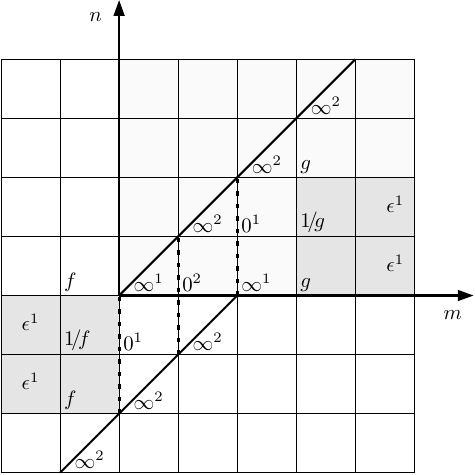}}
\smallskip\ \centerline{{Figure \four   }}
\medskip
This is shown in the lower left part of Figure \four. We find indeed that the products $x_{m-1,n} x_{m-1,n-1}$ and $x_{m-1,n-1}x_{m-1,n-2}$ are again of order $1+{\cal O}(\epsilon)$ and that $x_{m-1,n}$, $x_{m-1,n-1}$ and $x_{m-1,n-2}$ take finite, reciprocal, values at the limit $\epsilon\to0$. Thus what at the outset looked like an `arbitrary' initial condition, intended to mimic the $\{\infty^1,0^1,\infty^1\}$ lines of Figure \two, actually corresponds to an interaction of a line of $\infty^2$ with a double taishi $\epsilon^1, \epsilon^1$. The conclusion of this calculation then is that a triple of infinite lines of the form $\{\infty^1,0^2,\infty^1\}$ cannot exist. However, this does not mean that infinite lines of the form $\{\infty^p,0^q,\infty^r\}$ with $pqr\ne1$ {are impossible}. It can be verified that such lines do persist indefinitely if they satisfy the stability condition $\min[p,q,r]=q$.

In [\kdv] we presented a set of rules governing the dynamics of the interaction of one or more taishi with a diagonal line of infinities for the d-KdV equation. The basic assumption for this rule to work is that the line of infinities has weight 1/2. The result of an interaction with a line of infinities with general integer weight $q$ is thus obtained by repeating the procedure below $2q$ times. Citing [\kdv]:

\medskip
{\sl Starting sufficiently low,

- move upwards up to the first (i.e. lowest) horizontal strip in a taishi with non-zero weight and subtract 1 from its weight and add 1 to the weight of the strip just above it,

- then move to the next horizontal strip with non-zero weight above those two strips, subtract 1 from its weight and add 1 to the weight of the strip just above it,

- repeat the same procedure until there are no more taishi with strips with non-zero weight left.}
\medskip

No rule for the interaction of the taishi with a line of zeros was provided in [\kdv]. However the answer is simple. In order to treat such lines of zeros it suffices to follow the procedure described above while replacing `low' by `high', `upwards' by `downwards', 'subtract' by `add' and `above' by `below'. And of course one must keep in mind that an infinite line of zeros cannot exist unless it is flanked on both sides by lines of infinities, as shown in the bottom-right corner of Figure \two.

In the present paper we study the singularity structure of a lattice equation that is linked to the lattice KdV equation by a set of Miura transformations and which can therefore be regarded as a discrete modified KdV (mKdV) equation. We show that although there exist many similarities with the case of the lattice KdV equation described above, there are also essential differences. In particular the formulation of the symbolic representation of the dynamics for the taishi interactions in the case of mKdV is much simpler, allowing us to entertain hopes for a future rigorous explanation for their dynamics.
\bigskip
2. {\scap A discrete modified} KdV {\scap equation}
\medskip
The discrete form of the KdV equation was first derived by Hirota in [\hirota]. In his monumental work [\refdef\dagte] where he proposed the discrete analogue of the KP equation (an equation known since as the Hirota-Miwa [\refdef\miwa] equation) he also derived a discrete analogue of the modified KdV equation but only gave its bilinear form, i.e. in terms of tau functions. A nonlinear form of a potential discrete mKdV equation was proposed by Capel and collaborators in [\refdef\capel] and Hirota obtained in [\refdef\hirom] the nonlinear form of the mKdV he had previously derived. As some of the present authors showed in [\refdef\scimi] this equation in fact coincides with that of Capel et al., after some elementary transformations. 

In [\refdef\levi] Levi and Yamilov derived what at the time seemed to be two new partial difference equations. They obtained these equations from the combination of Miura transformations linking the differential-difference Volterra equation to a modified version thereof. They constructed the Lax pairs for both equations and concluded that these systems are integrable. The two equations of Levi and Yamilov were analysed in [\refdef\notso] by some of the present authors. We started by obtaining their elementary singularities and derived their bilinear forms. The singularity patterns we obtained were all confined (thus strengthening the argument in favour of integrability) but the bilinear forms turned out to coincide with those already known for the discrete KdV and mKdV equations. Proceeding to the continuum limit we showed that the two equations of Levi and Yamilov are indeed discrete analogues of the KdV and mKdV equations. The fact that different discrete equations can have the same continuum limit is not astonishing. The discrete integrable domain is far richer than the continuous one and when taking a continuum limit of an integrable discrete system to a continuous system, there are simply not that many possibilities for the ensuing continuous integrable system. Unfortunately, this also means that naming discrete systems based on their continuum limit is not quite appropriate, but a better option is not always available. In what follows we shall work with the discrete modified KdV equation proposed by Levi and Yamilov in [\levi] and, for the sake of brevity, we shall refer to it as the d-mKdV equation (hoping that there will be no confusion with the Hirota-Capel lattice equation).

The discrete modified KdV equation we shall work with has the form 
$$(1+v_{m,n}v_{m+1,n})(k v_{m+1,n+1}+v_{m,n+1}/k)=(1+v_{m,n+1}v_{m+1,n+1})(k v_{m,n}+v_{m+1,n}/k),\eqdef\ztri$$
where $k$ is a non-zero constant such that $k^4\neq1$ (lest the equation become degenerate, as pointed out in [\levi]).
We remark that, given the form of (\ztri), the equation is invariant under a sign change of either $v$ or $k$.
An equivalent form of (\ztri) can be obtained by rearranging the terms on both sides:
$$(1-v_{m,n}v_{m,n+1})(k v_{m+1,n+1}-v_{m+1,n}/k)=(1-v_{m+1,n}v_{m+1,n+1})(k v_{m,n}-v_{m,n+1}/k).\eqdef\ztes$$
Equation (\ztri) being a discrete version of the modified KdV equation, it is expected to be related to the d-KdV equation by a Miura transformation. Curiously, this relation was not given in [\notso] and we remedy this here. The Miura relations have the form
$$x_{m,n}x_{m+1,n}={(k-v_{m,n})(k+v_{m+1,n})\over(kv_{m,n}-1)(kv_{m+1,n}+1)},\eqdef\zpen$$

$$x_{m,n}x_{m,n+1}={(k-v_{m,n})(k+v_{m,n+1})\over(kv_{m,n}+1)(kv_{m,n+1}-1)}.\eqdef\zhex$$

Eliminating $v$ between these two relations gives relation (\zena) for $x$ and, similarly, the elimination of $x$ from (\zpen) and (\zhex) leads to the d-mKdV equation (\ztri) above. To obtain equation (\ztes) by Miura transformation from the d-KdV equation, it suffices to change each occurrence of $k$ (or $v$) in (\zpen) and (\zhex) to $-k$ (or $-v$).

In the following, all results we shall present are obtained from the evolution under equation (\ztri)  in the positive $m$ and $n$ directions -- i.e. towards the NE on the lattice, as indicated in Figure \figdef\five\ -- of specific sets of initial conditions given on a staircase.

The elementary singularity of d-mKdV was already obtained in [\notso]. Whenever, due to a special choice of initial conditions, $v$ takes the value $k$
at a point $(m,n)$ on the lattice, the values $1/k$ and $-1/k$ appear at points $(m,n+1)$ and $(m+1,n)$, respectively, whereas $v_{m+1,n+1}$ takes the value $-k$ and the singularity disappears. It is confined.
Figure \five\ below shows such a singularity pattern.
\bigskip
\centerline{\includegraphics[width=5.5cm,keepaspectratio]{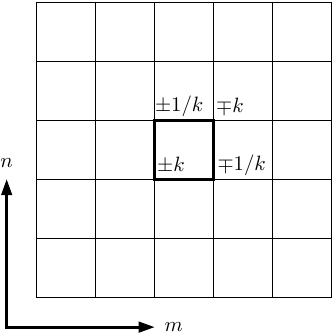}}
\smallskip\quad\centerline{{Figure \five   }}
\medskip
When one compares this singularity pattern with the simplest one for d-KdV, presented in Figure \one, it is tempting to identify a 0 in the d-KdV case with a value $\pm k$ for d-mKdV, and similarly an $\infty$ for d-KdV with a $\pm1/k$ in the present case. The results presented in this paper sustain this analogy, up to a point, but as we shall explain this is by no means a strict equivalence.

In the light of the results of [\kdv] and the comments just above, we must also look for other possible singularities. It turns out that an infinite line {of values $1/k$, running from SW to NE, can exist.} It plays the same role as a line of infinities for d-KdV. Figure \figdef\six\ shows such a singularity.
\bigskip
\centerline{\includegraphics[width=5.0cm,keepaspectratio]{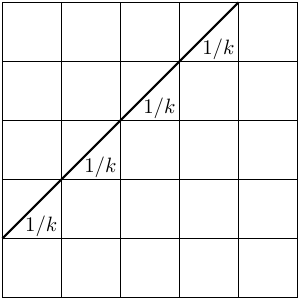}}
\smallskip\ \centerline{{Figure \six   }}
\medskip
It goes without saying that, given the symmetry of (\ztri), an infinite line of values $-1/k$ is also a possibility. In fact, several infinite lines can coexist, and can even be adjacent, without any restrictions on the plus or minus sign associated to each line. 

The interesting question at this point, given the capital role they play in the case of the d-KdV equation, is whether taishi-like singularities also exist for the d-mKdV equation (\ztri). It turns out that this is indeed the case. Already by direct  inspection of (\ztri) one sees that if $v_{m,n}v_{m+1,n}=-1$, for some value of $n$, the same relation automatically holds also at $n+1$. Similarly, from (\ztes), if $v_{m,n}v_{m,n+1}=1$ for some $m$ the same is true at $m+1$. (To be fair, the reason we chose to work with the Levi-Yamilov form of d-mKdV was precisely the fact that the taishi condition can be assessed by simple inspection of the equation). Figure \figdef\seven\ shows a vertical and a horizontal taishi.
\bigskip
\centerline{\includegraphics[width=6cm,keepaspectratio]{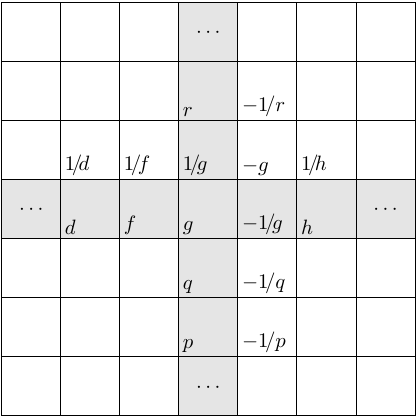}}
\smallskip\ \centerline{{Figure \seven   }}
\medskip
It is also interesting to point out that the taishi conditions in the present case can be obtained from those for the d-KdV equation through the Miura transformations. Assuming for instance that $x_{m,n}x_{m+1,n}=1$ in (\zpen), one readily obtains the condition $v_{m,n}v_{m+1,n}=1$, and similarly in the other direction. 

Having presented the elementary singularities of d-mKdV, we now turn to the study of more complex singularity patterns for which we shall see that the above `simple' correspondence between the singularities for the d-KdV and d-mKdV equations breaks down.

In what follows, when we say that a singularity occurs at $k$, we mean that the corresponding calculations are actually performed with the value $k+\gamma \epsilon$ where $\gamma$ is a factor specific to each singularity (and similarly for singularities at $1/k$ as well as for the taishi conditions).
\bigskip
3. {\scap The interaction of singularities of the discrete mKdV equation}
\medskip
Before addressing the question of interaction of the taishi with the other singularities of the discrete mKdV equation,  it is interesting to show the singularity pattern generated by two adjacent values $v_{m,n+1}$ and $v_{m+1,n}$ that are equal to $k$. The resulting confined singularity pattern is shown in Figure \figdef\eight.
\bigskip
\centerline{\includegraphics[width=5.5cm,keepaspectratio]{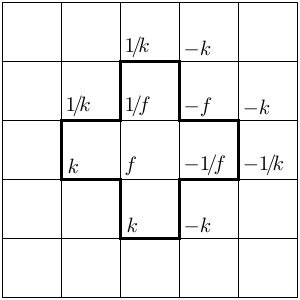}}
\smallskip\ \centerline{{Figure \eight   }}
\medskip
Comparing this figure to Figure \one, obtained in the case of KdV, we remark that while the values on the outside corners of the singularity pattern are, as expected, a value $k$ where we would have had a 0 for d-KdV and a value $1/k$ where an infinity would have appeared for d-KdV, the singularities appearing in the interior corners of the cross figure take neither of these two special values {and the analogy between the singular values for these two systems clearly breaks down in this case}. Still, the pattern on the interior corners does follow the inversions and changes of sign of the elementary singularity of d-mKdV, just as is the case for the d-KdV. (Note that a pattern similar to that of Figure \eight\ can also be obtained in the case of d-KdV, but only when the `epsilons' in the two starting singularities are taken to be strictly identical).

The simplest taishi interaction one can think of is one where the taishi impinges upon an elementary singularity, either head-on or tangentially. Figure \figdef\nine, obtained for initial conditions on a staircase that includes the values $1/d$, $d$ and $k$, shows a head-on collision. 
\bigskip
\centerline{\includegraphics[width=5.5cm,keepaspectratio]{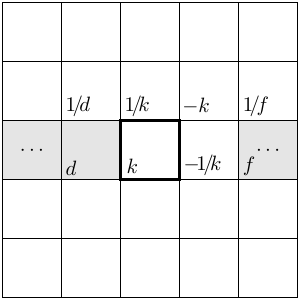}}
\smallskip\ \centerline{{Figure \nine   }}
\medskip
We remark that the elementary singularity is not perturbed by its encounter with the taishi. On the other hand if the collision is off-centre but still touches the elementary singularity pattern tangentially on its north side, as in the case shown in Figure \figdef\ten, a mirror-image of the singularity is created above (i.e. to the north of) the original singularity pattern. The taishi however exits the interaction region completely unperturbed. Note that Figure \ten\ was obtained from initial conditions on a staircase that includes the values $1/d$, $d$ and the value $k$ to the SE of the value $d$, but none of the other singular values.
\bigskip
\centerline{\includegraphics[width=5.5cm,keepaspectratio]{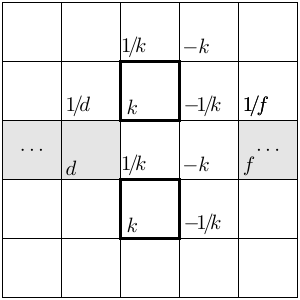}}
\smallskip\ \centerline{{Figure \ten   }}
\medskip
Similar behaviours in the case of the taishi interacting with the elementary singularity were observed for the d-KdV equation.

Next we turn to the interaction of a taishi with an infinite line of values $1/k$. The result of such an interaction is presented in Figure \figdef\eleven, which was obtained from initial conditions on a staircase that includes the values $1/f$, $f$ and one value $1/k$ either to the bottom or to the right of $f$.
\bigskip
\centerline{\includegraphics[width=5.5cm,keepaspectratio]{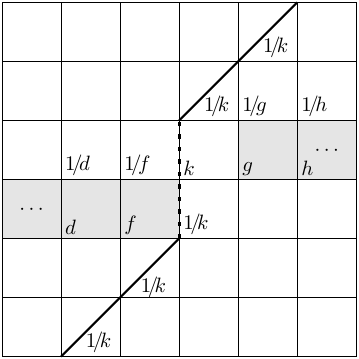}}
\smallskip\ \centerline{{Figure \eleven   }}
\medskip
We remark that the result of this interaction is a shift of both the taishi as well as of the infinite line. This situation is reminiscent of what happens in the case of KdV, but the details of the interaction are quite different. In the case of the d-KdV equation the interaction of a weight 1 taishi with a line of infinities with that same weight results in an outgoing taishi shifted upwards by two positions, whereas we only observe a single shift in the present case.

In order to analyse these interactions in more detail we must relinquish the tacit assumption that the difference from the value 1 for the product of adjacent values in the case of the taishi, and from the value $1/k$ for the infinite line, are both of first order in the small parameter $\epsilon$. As explained in the introduction, we must introduce weights for both the taishi and the infinite line. For lack of a more convenient notation we shall use, in the case of $1/k$, a subscript to indicate the order in $\epsilon$ of the perturbation of the value $1/k$ for $v$ . For instance, when we use the symbol $1/k_2$ we mean this particular $v$ takes the value $1/k+{\cal O}(\epsilon^2)$. When no subscript is given, this is to be understood as simply meaning $1/k+{\cal O}(\epsilon)$.

We first show an interaction of a taishi of weight 2 with an infinite line of weight 1 (Figure \figdef\twelve). 
\bigskip
\centerline{\includegraphics[width=6cm,keepaspectratio]{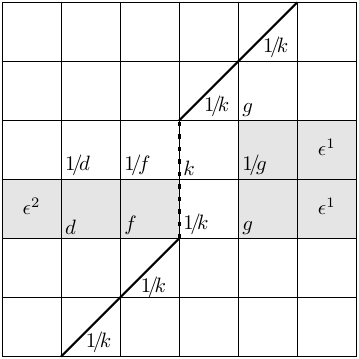}}
\smallskip\ \centerline{{Figure \twelve   }}
\medskip
We remark that the taishi is split into two parts, sharing equally the initial weights, which is again quite different from the d-KdV case where the impinging taishi would not split but would just be shifted upwards by one position. 

The situation when a taishi of weight 1 impinges upon an infinite line of weight 2 is shown in Figure \figdef\thirteen. 
\bigskip
\centerline{\includegraphics[width=6cm,keepaspectratio]{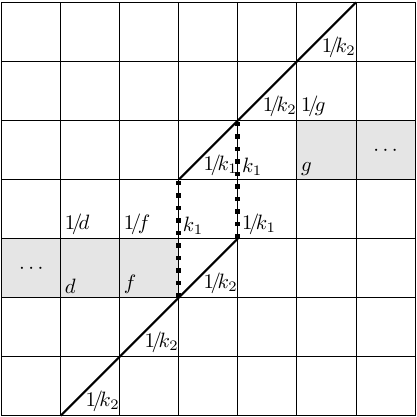}}
\smallskip\ \centerline{{Figure \thirteen   }}
\medskip

We could of course continue presenting more and more examples of taishi interactions. However, the set of examples we have at our disposal already suffices to formulate a symbolic representation of the dynamics that describes the interaction of a taishi with an infinite line of $\pm 1/k$. This will be the subject of  section 4. 

Before proceeding any further, however, it is interesting to present yet one more case where the analogy between the d-KdV and the d-mKdV equations breaks down.  As shown in Figure \two\ the d-KdV equation possesses a singularity pattern consisting in three infinite lines where a zero is interspersed between two infinities. (And of course also more complicated patterns where infinities alternate with zeros). As we saw in the previous section, there is a strong resemblance between the d-KdV and d-mKdV singularities where, in the latter, values $1/k$ and $k$ play the roles played by $\infty$ and 0 for the former equation. One would thus be tempted to posit the existence of a triple of infinite lines of singularities for d-mKdV, where a value of $k$ is interspersed between two $1/k$ values. In order to investigate this possibility we start with precisely such an initial condition, namely $v_{m-1,n}=1/k, v_{m,n}=k$ and $v_{m+1,n}=1/k$. (The weights of all three are taken to be equal to 1, i.e. $v_{m-1,n}=1/k+{\cal O}(\epsilon)$ etc.; we have verified that choosing different weights does not alter the conclusions). The result of the evolution of this specific initial condition is shown in the Figure \figdef\fourteen\ below. 
\bigskip
\centerline{\includegraphics[width=6cm,keepaspectratio]{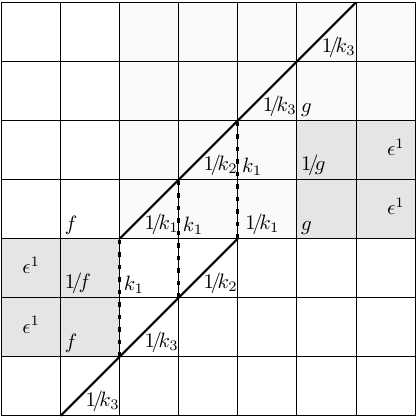}}
\smallskip\ \centerline{{Figure \fourteen   }}
\medskip
We observe that the initial condition $1/k, k,1/k$ does not survive. Looking at the upper right (slightly greyed) corner of the figure, one sees that only a singularity $1/k$ with weight 3 emerges, accompanied by a taishi of overall weight 2. In fact, iterating backwards one sees that the initial condition we introduced was just the result of the interaction of a taishi of weight 2 with an infinite line $1/k$ of weight 3. This situation is of course strongly reminiscent of the instability observed in the case of Figure \four, for d-KdV, due to the choice of the weight of the zero in the middle of the initial condition. In the case of the d-KdV equation however there exist infinitely many choices of the weight of this zero that {do} lead to a stable singularity pattern. It seems that this is not possible for the d-mKdV equation. The analogy between the singularity patterns of the two equations is thus less straightforward than one might have surmised. 
\bigskip
4. {\scap A symbolic representation of the dynamics for the taishi interactions}
\medskip
In the previous section we have presented several examples of interactions of a taishi with an infinite line of $1/k$. While in these relatively simple examples it was easy to follow the evolution of the taishi, this is by no means always the case. We therefore need a rule that will allow us to calculate the shape of the taishi after it emerges from the interaction region, as well as the shift of the infinite line of $1/k$ due to the interaction. The answer to the latter question is simple: every single interaction with a taishi -- be it simple or double -- results in an upwards shift of the infinite line by two lattice steps. 

The changes the interaction induces on the d-mKdV taishi themselves can be described using a symbolic representation similar to that for the d-KdV case, though slightly simpler. In the case of the d-KdV equation the symbolic representation of the dynamics for its taishi was formulated assuming a somewhat artifical weight of 1/2 for the line of infinities, requiring extra intermediary steps which are now no longer necessary. As we shall see, the rules for the symbolic representation of the dynamics for the d-mKdV taishi can be formulated based on the natural assumption that the infinite line it interacts with has weight 1. Keeping this essential difference in mind we can now formulate the symbolic rules in exactly the same terms as for the d-KdV equation:

\medskip
{\sl Starting well-below the lowest taishi (in the $n$ direction),

- move upwards up in $n$ to the first (i.e. lowest) horizontal strip in a taishi with non-zero weight and subtract 1 from its weight and add 1 to the weight of the strip just above it,

- then move to the next horizontal strip with non-zero weight above those two strips (in the $n$ direction), subtract 1 from its weight and add 1 to the weight of the strip just above it,

- repeat the same procedure until there are no more taishi with strips with non-zero weight left.}
\medskip

This is the rule for the interaction of an arbitrary collection of taishi with an infinite line of $1/k$ with weight 1. If the weight $q$ of the line of $1/k$ is greater than 1, then the above procedure should simply be repeated $q$ times.

The interaction rule for a collection of taishi with a weight 1 line of values $1/k$ can also {be rephrased in the language of so-called `Box\&Ball systems' [\refdef\tamatsu]}.
Let us represent the weights of all the taishi, before interaction, as entries in an infinite vector $U$ parallel to the $n$ axis on the lattice (the vector $U$ only contains the weights of the taishi at each position in $n$, before interaction, and neglects all weights associated to any infinite lines the taishi will interact with) and think of the changes in weights described by the above rule, as the result of an interaction with a carrier that moves along the $n$ axis, from position $-\infty$ to $+\infty$, interacting with the entries in $U$ at each successive position it passes through. This carrier has maximal capacity 1, i.e. it can only take the values 0 and 1 (in which case we say that it is empty or full, respectively) and it is always empty for sufficiently negative positions in $n$. 
An empty carrier does not interact with zero entries in the vector $U$ but if the carrier arrives empty at a position in $U$ with a non-zero entry, then the value in $U$ at this position decreases by one and the value of the carrier is updated to become 1. On the other hand, a carrier that arrives full at a certain position decreases its state to 0 and adds 1 to the value in $U$ at that position. {Note that this} necessarily happens at each position just above a position where the carrier was updated from empty to full. {Hence, the above interaction rule can be interpreted as the update rule for a Box\&Ball system with carrier [\tamatsu], where $U$ represents an infinite array of boxes that can each contain an arbitrary number of balls (the weights of the taishi) and where the carrier, $V$, has capacity 1.}

If we denote the updates of the values of the carrier by $\hat V$ and the {(local)} update of the vector $U$ by $\bar U$, then the interaction rule at each position $n$ in the vector $U$ can be expressed as
$$V=1\quad \Rightarrow\quad \hat V = 0,\quad \bar U= U + 1,\hskip1.5cm\eqdef\znana$$
$$V=0\quad \Rightarrow\quad \hat V = \min[1,U],\quad \bar U= U + V - \hat V,\eqdef\zhachi$$
which can be summarized in the closed form
$$\hat V= \min[U, 1-V]\,,\quad\bar U= U + V - \hat V,\eqdef\zkyu$$
under the boundary conditions $V,U=0$ for $|n|\sim\infty$ (for a vector $U$ that only has non-negative integer entries). {Readers familiar with Box\&Ball systems will immediately notice that the update rule (\zkyu), subject to said boundary conditions, is in fact the same as that of the famous Takahashi-Satsuma soliton cellular automaton [\refdef\takasa]. We shall come back to this point in section 5.}

Using this rule, the dynamics corresponding to figures \eleven, \twelve\ and \thirteen\ can now be simply represented by the diagrams below:

$$\matrix{ 0\cr 0\cr 0\cr \bf1\cr 0\cr 0}\Longrightarrow\matrix{0\cr0\cr \bf1\cr0\cr0\cr0}\hskip2.5cm\matrix{ 0\cr 0\cr 0\cr \bf2\cr 0\cr 0}\Longrightarrow\matrix{0\cr0\cr \bf1\cr \bf1\cr0\cr0}\hskip2.5cm\matrix{ 0\cr 0\cr 0\cr \bf1\cr 0\cr 0}\Longrightarrow\matrix{0\cr0\cr1\cr0\cr0\cr0}\Longrightarrow\matrix{0\cr \bf1\cr0\cr0\cr0\cr0}$$

in which only the (immediately relevant) entries of the vector $U$ are given, and where numbers in boldface correspond to the weights in the taishi before interaction (on the left) and after it has left the interaction region (on the right).

The interaction shown in Figure \fourteen\ however requires some explanation. First, let us consider the generic interaction of the taishi of Figure \fourteen\ with an infinite line of weight 3, i.e. obtained for the evolution of initial conditions (on a staircase) that include the values $1/f, f$ and $1/k^3$ (as shown in Figure \figdef\fifteen) but none of the other singular values.
\bigskip
\centerline{\includegraphics[width=6cm,keepaspectratio]{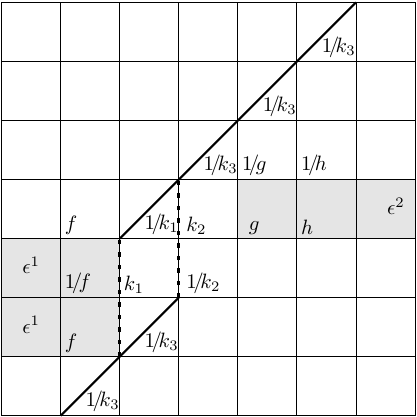}}
\smallskip\ \centerline{{Figure \fifteen   }}
\medskip
The corresponding symbolic representation is given by:
$$\matrix{ 0\cr 0\cr 0\cr \bf1\cr \bf1\cr 0}\Longrightarrow\matrix{0\cr0\cr0\cr2\cr0\cr0}\Longrightarrow\matrix{0\cr0\cr1\cr1\cr0\cr0}
\Longrightarrow\matrix{0\cr 0\cr \bf2\cr 0\cr0\cr0}$$
where the numbers in boldface again represent the initial and final weights of the impinging taishi.
The interaction depicted above is of course clearly different from that in Figure \fourteen.
However, one should not forget that the initial conditions that lead to the interactions shown in Figure \fourteen\ were chosen such as to have consecutive values $1/k, k$ and $1/k$ in the $m$ direction, instead of the values that would naturally arise from the taishi interaction, namely $1/k, k_2,g,h,\cdots$ (as in Figure \fifteen). 
This means that we have, in fact, {enforced one extra interaction with what turns out to be a diagonal with weight 1 at the point of interaction (as shown by the thick line in Figure \figdef\sixteen\ below).}
\bigskip
\centerline{\includegraphics[width=6cm,keepaspectratio]{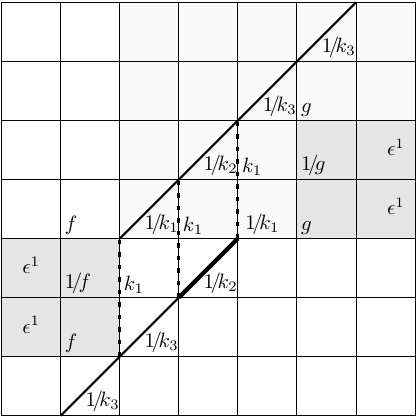}}
\smallskip\ \centerline{{Figure \sixteen   }}
\medskip
The presence of this (weight 1) line alters the symbolic representation to 
$$\matrix{ 0\cr 0\cr 0\cr \bf1\cr \bf1\cr 0}\Longrightarrow\matrix{0\cr0\cr0\cr2\cr0\cr0}\Longrightarrow\matrix{0\cr0\cr1\cr1\cr0\cr0}
\Longrightarrow\matrix{0\cr 0\cr 2\cr 0\cr0\cr0}\Longrightarrow\matrix{0\cr \bf1\cr \bf1\cr 0\cr0\cr0}$$
in perfect agreement with what was obtained in Figure \fourteen. 

Imposing special (singular) values $k$ or $1/k$ at specific lattice sites can have a deep effect on the evolution, transforming the singularity patterns one would otherwise obtain for less specific initial values. Suppose for instance that in the cross pattern shown in Figure \eight\ we demand that the value $f$ appearing in the interior of the cross is exactly equal to $1/k$, in analogy to what happens in the case of the d-KdV equation. The consequence of this choice is a profound modification of the singularity pattern, as can be seen in Figure \figdef\seventeen. 

The value $1/k$ we imposed instead of $f$ requires -- thinking of the backward evolution -- the existence of a whole line of $1/k$ interacting with a taishi of weight 1 in the SW direction. As a result, this taishi is shifted upwards once and exits the interaction area as expected. The singularities around and inside the cross follow the same pattern as in d-KdV, {\sl mutatis mutandis}, but the exiting taishi modifies the righthand side of the cross. So while a parallel with d-KdV does exist up to a certain point, treating the value $k$ in d-mKdV as the strict analogue of the d-KdV zero  does not seem to work, apart from the simplest cases. This makes the study of the singularities of d-mKdV all the more interesting.

\bigskip
\centerline{\includegraphics[width=5.5cm,keepaspectratio]{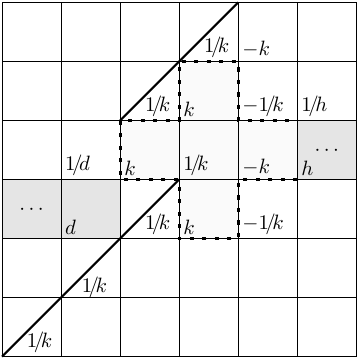}}
\smallskip
\ \centerline{{Figure \seventeen   }}
\medskip
5. {\scap Conclusion and outlook}
\medskip
In this paper we studied the singularities of the discrete modified KdV equation and their interactions. This study was motivated by an exhaustive investigation of the possible singularities for the discrete KdV equation that we carried out recently [\kdv] and which led to the identification of three different types of singularities. 
The first type are localised, `confined', singularities. These were already identified in [\sincon] and their discovery led to the formulation of the hypothesis that integrable systems possess confined singularities. (It was discovered later that this property does not necessarily hold for systems integrable through linearisation, although when the integrability is obtained by spectral methods no violation of the singularity confinement property is known).  The second type consists of singularities with infinite extent,  more precisely of diagonals of infinities going in the south-west to north-east direction on the lattice and, in some cases,  of lines of infinities alternating with lines of zeros. We must stress here again that the infinite extent of such a singularity is {\sl not} incompatible with integrability. 

The most interesting singularities however are those of the third type, the existence of which was completely unexpected. These singularities consist of horizontal (or vertical) strips, single or multiple, in which the product of two vertically (horizontally) adjacent values is 1 (or $-1$). The particularity of this third type of singularity is that it can interact with singularities of all the other types, leading to very rich singular behaviour for the d-KdV equation. We decided to introduce a new name for this type of singularity, opting for the term `taishi', inspired by the Japanese word for `band'. 

In view of these results the question that of course immediately arises is whether these taishi are an anomaly. Is their existence specific to the d-KdV equation or, might they be a feature that one should expect from other integrable lattice equations as well? It was thus natural to focus on the best candidate to study this problem, given its affinity to KdV:  its modified counterpart. For our study we chose a particular discrete form of the modified KdV equation, first proposed by Levi and Yamilov [\levi]. The main advantage of this choice was the fact that the existence of the taishi could be assessed at a glance, given the form of the equation. It thus turned out that this d-mKdV indeed possesses the same type of singularities as the d-KdV equation (with the exception of the infinite lines of zeros flanked by infinite lines of infinities which do not exist in the d-mKdV case). Moreover, the taishi still interact with the other types of singularities, in ways reminiscent of the d-KdV case. 
However, `reminiscent' does not mean `identical' and several key aspects of the interactions for the d-mKdV singularities differ from those for the d-KdV equation. In particular, the symbolic representation of the dynamics that describes the interaction of a set of taishi with an inifinite line of singular values turns out to be simpler than for the d-KdV equation. 

As explained in section 4, the symbolic representation of the interaction of a (finite) set of taishi with an infinite line of singular values $\pm1/k$ with weight $q$, is in fact equivalent to $q$ time steps in a Box\&Ball system with carrier [\tamatsu], with infinite capacity for the boxes (the number of balls in which represent the weigths of the taishi) and with a carrier that can only carry 1 ball. This particular system is the special case $L=+\infty$, $M=1$ of the system
$$U_j^{t+1} = V_j^t + \max[0, U_j^t+V_j^t-M] - \max[0,U_j^t+V_j^t-L],\eqdef\zju$$
$$V_{j+1}^{t} = V_j^t + U_j^t-U_j^{t+1},\eqdef\zjuichi$$
where $U_j^t$ represents the number of balls contained at time $t$ in the box with index $j$ and where $V_j^t$ represents the number of balls in the carrier when, at time $t$, it arrives at box $j$. Under the boundary conditions $U_j^t, V_j^t=0$ for large enough $|j|$ (at all $t$) and for non-negative integer initial conditions $U_j^0$, this system describes the evolution of a finite number of balls ($\sum_{j=-\infty}^{+\infty}U_j^0$) over an  infinite array of boxes -- each with fixed capacity $L$ -- due to the movement of a carrier with capacity $M$ in the (positive) $j$ direction [\tamatsu]. A simple calculation shows that the particular case $L=+\infty$, $M=1$ can also be written as (\zkyu), which under the same boundary conditions actually corresponds to the rule that defines one time step in the Takahashi-Satsuma soliton cellular automaton [\takasa], but in this case for the carrier variable $V$ and not for the $U$ variable that represents the boxes. This correspondence between two apparently very different systems is a simple consequence of the self-duality of the system $(\zju,\zjuichi)$ under the involution $(U,V;L,M;t,j)\to(V,U;M,L;j,t)$.

More important however is that the system $(\zju,\zjuichi)$ is, in fact, an ultradiscrete (or tropical) version of a (so-called Harrison-type) Yang-Baxter map given in [\refdef\harri], which is known to define the Hirota-Capel discrete version of the modified KdV equation [\refdef\knw]. We therefore arrive at the striking conclusion that the symbolic representation of the interaction of the taishi with type-two singularities for our d-mKdV equation (and,  by extension, also for the d-KdV equation) is actually equivalent to a tropical Yang-Baxter map (or combinatorial $R$-matrix [\refdef\symmten]) that governs the {\sl algebraic} integrability of the ultradiscrete version of the mKdV equation (or, depending on your point of view, of the quantum integrable system with the same $A_1^{(1)}$ symmetry, obtained through crystallization [\refdef\inoue]). The fact that soliton interactions define Yang-Baxter maps has been known for quite a while [\refdef\veselov,\refdef\goncha] but that interactions of singularities are similarly related is quite remarkable.

The results presented in the present paper thus not only confirm that the behaviour we observed in the KdV case was not a singular occurrence (no pun intended), but also point at a possibly deep connection between analytic aspects (singularities) and algebraic aspects (Yang-Baxter maps) of integrability in lattice equations. Besides the exploration of this intriguing connection, we are also naturally led to speculate upon other possible paths for future studies. 
Singularities in the solutions of multidimensionally consistent lattice type-Q equations in the ABS-classification have been investigated in [\refdef\atkins], but singularities in non type-Q equations, like those studied here, present particular difficulties.

The first question is whether the singularity structures observed here also exist in other integrable lattice equations, and under which guises. After all, KdV and m-KdV are like siblings and observing similar behaviour in both is not all that astonishing. 
But even restricting ourselves to the KdV case, the fact that the symbolic representation of the dynamics of the d-mKdV singularities are very simple raises the hope of possibly finding a rigorous proof for their formulation. In both [\kdv] and the present paper the symbolic rules were deduced by the observation of a slew of cases, but a rigorous proof might also bring us closer to understanding the link to Box\&Ball systems (possibly a very tall order). 
A different direction one could consider is that of establishing singular rules for these equations. Singular rules have been introduced in numerical analysis [\refdef\wynn,\refdef\brez] and, in particular, in convergence accelerators in order to deal with accidentally vanishing denominators. The existence of such rules is intimately related to the singularity confinement property. It would thus be interesting to study the possible existence of such rules for the various singularities we have discovered so far. 

\bigskip
{\scap Acknowledgements}
\medskip
RW would like to acknowledge support from the Japan Society for the Promotion of Science (JSPS), through JSPS grant
number 18K03355. He also wishes to thank Y. Nakata for useful discussions on the Box\&Ball system of section 4.
\bigskip
{\scap References}
\medskip
\begin{description}
\item{[\goriely]} A. Goriely, `Integrability and Nonintegrability of Dynamical Systems', World Scientific, Singapore (2001).
\item{[\physrep]} A. Ramani, B. Grammaticos and T. Bountis, Physics Reports {\bf 180} (1989) 159.
\item{[\ars]} M.J. Ablowitz, A. Ramani and H. Segur, Lett. Nuov. Cim. {\bf 23} (1978) 333.
\item{[\sincon]} B. Grammaticos, A. Ramani and V. Papageorgiou, Phys. Rev. Lett. {\bf 67} (1991) 1825.
\item{[\hirota]} R. Hirota, J. Phys. Soc. Japan {\bf 43} (1977) 1424.
\item{[\doyong]} D. Um, R. Willox, B. Grammaticos and A. Ramani, J. Phys. A: Math. Theor. {\bf 53} (2020) 114001.
\item{[\kdv]} D. Um, A. Ramani, B. Grammaticos, R. Willox and J. Satsuma, J. Phys. A: Math. Theor. {\bf 54} (2021) 095201.
\item{[\dagte]} R. Hirota, J. Phys. Soc. Japan {\bf 503} (1981) 785.
\item{[\miwa]} T. Miwa, Proc. Japan Acad. {\bf 58} (1982) 9.
\item{[\capel]} H. Capel, F.W. Nijhoff and V. Papageorgiou  Phys. Lett A {\bf 153} (1991) 377.
\item{[\hirom]} R. Hirota, J. Phys. Soc. Japan {\bf 672} (1998) 234.
\item{[\scimi]} B. Grammaticos, A. Ramani, C. Scimiterna and R. Willox, J. Phys. A: Math. Theor. {\bf 44} (2011) 152004
\item{[\levi]}  D. Levi and R.I. Yamilov, ``On a nonlinear integrable difference equation on the square 3D-inconsistent'',
 preprint (2009) arXiv:0902.2126 [nlin.SI].
\item{[\notso]} A. Ramani, B. Grammaticos, R. Willox and J. Satsuma, J. Phys. A: Math. Theor.  {\bf 42} (2009) FT 282002.
\item{[\tamatsu]} D. Takahashi and J. Matsukidaira, J. Phys. A: Math. Gen. {\bf 30} (1997) L733.
\item{[\takasa]} D. Takahashi and J. Satsuma, J. Phys. Soc. Japan {\bf 59} (1990) 3514.
\item{[\harri]} V. G. Papageorgiou, A. G. Tongas and A. P. Veselov, J. Math. Phys. {\bf 47} (2006), 083502.
\item{[\knw]} S. Kakei, J.J.C. Nimmo and R. Willox, Glasgow Math. J. {\bf 51A} (2009) 107.
\item{[\symmten]} G. Hatayama, K. Hikami, R. Inoue, A. Kuniba, T. Takagi and T. Tokihiro, J. Math. Phys. {\bf 42} (2001) 274.
\item{[\inoue]} R. Inoue, A. Kuniba and T. Takagi, J. Phys. A: Math. Theor. {\bf 45} (2012) 073001.
\item{[\veselov]} A.P. Veselov, Phys. Lett. A {\bf 314} (2003) 214.
\item{[\goncha]} V.M. Goncharenko and A.P. Veselov, ``Yang-Baxter maps and matrix solitons'', in `New Trends in Integrability and Partial Solvability', A.B. Shabat et al. (eds.), Springer, Dordrecht (2004) pp. 191.
\item{[\atkins]} J. Atkinson, SIGMA {\bf 7} (2011) 73.
\item{[\wynn]} P. Wynn, BIT {\bf 3} (1963), 175.
\item{[\brez]} C. Brezinski, `Algorithmes d'acc\'el\'eration de la convergence: \'etude num\'erique', Technip, Paris (1978).
\end{description}

\end{document}